\documentclass[12pt]{iopart}
\usepackage{iopams,graphicx}

\usepackage{lmodern}
\usepackage[T1]{fontenc}
\usepackage{bbold}
\graphicspath{{Graphics/}}
\usepackage{xcolor}

\usepackage{soul}
\usepackage[mathscr]{euscript}
\usepackage{cite}
\usepackage[super]{nth}
\usepackage{float}  
\usepackage{comment}   
\usepackage[colorlinks, citecolor = blue, urlcolor = blue]{hyperref}

\begin{document}

\title{Diffusion with preferential relocation in a confining potential}
\author{Denis Boyer}
\address{Instituto de F\'\i sica, Universidad Nacional Aut\'onoma de M\'exico,
Ciudad de M\'exico 04510, M\'exico}
\author{Martin R. Evans}
\address{SUPA, School of Physics and Astronomy, University of Edinburgh, 
Peter Guthrie Tait Road, Edinburgh EH9 3FD, United Kingdom}
\author{Satya N. Majumdar}
\address{LPTMS, CNRS, Universit\'e Paris-Saclay,  91405 Orsay,  France}

\begin{abstract}
We study the relaxation of a diffusive particle confined in an arbitrary external potential and subject to a non-Markovian resetting protocol. With a constant rate $r$, a previous time $\tau$ between the initial time and the present time $t$ is chosen from a given probability distribution $K(\tau,t)$, and the particle is reset to the position that it occupied at time $\tau$. Depending on the shape of $K(\tau,t)$, the particle either relaxes toward the Gibbs-Boltzmann distribution or toward a non-trivial stationary distribution that breaks ergodicity and depends on the initial position and the resetting protocol. From a general asymptotic theory, we find that if the kernel $K(\tau,t)$ is sufficiently localized near 
$\tau=0$, i.e., mostly the initial part of the trajectory is remembered and revisited, the steady state is non-Gibbs-Boltzmann. Conversely, if $K(\tau,t)$ decays slowly enough or increases  with $\tau$, i.e., recent positions are more likely to be revisited, the probability distribution of the particle tends toward the Gibbs-Boltzmann state at large times. In the latter case, however, the temporal approach to the stationary state is generally anomalously slow, following for instance an inverse power law or a stretched exponential, if $K(\tau,t)$ is not too strongly peaked at the current time $t$.  These findings are verified by the analysis of several exactly solvable cases and by numerical simulations.
\end{abstract}

\maketitle

\section{Introduction}
The effect of memory on a diffusive process is a general problem of
wide importance (see e.g. \cite{Stanislavsky00,BB2008,MSVV2013}) and it is known that memory can significantly
change the fundamental properties of diffusion. Non-Markov processes commonly give rise to anomalous diffusion and very slow relaxation toward equilibrium \cite{MK2000,EL2018}, and even to non-ergodic behaviour, where long-time average quantities of individual particles along their trajectory strongly deviate from the Gibbs-Boltzmann equilibrium \cite{B1992,BB2005}. Systems with global memory, where the transition probabilities depend on the whole previous history, are particularly prone to ergodicity breaking and localisation phenomena at late times \cite{B2016,FCBGM2017}. 

A simple model introduced in \cite{BEM2017} is the preferential
relocation model (sometimes referred to as `resetting with memory' or
`the monkey walk'). At the microscopic level,  in addition to nearest
neighbour random walk dynamics, the  diffusive particle may return
with a resetting rate $r$ to
a site where it was present at a previous time $\tau$, with $0 \leq \tau \leq t$ and
where $t$ denotes the present time. The preferential
relocation model selects this previous  time $\tau$ according to a distribution $K(\tau,t)$ which defines the memory kernel.
If  $K(\tau,t) = \delta(\tau)$ the model reduces to stochastic
resetting to the initial condition \cite{EM2011}. Whereas if
$K(\tau,t) = \delta(\tau-t)$
there is no memory and simple diffusion is recovered. The case
$K(\tau,t) = 1/t$  corresponds to selecting $\tau$ uniformly from the past and has been studied in the context of animal mobility \cite{BS2014,BR2014,MPCM2019}. In this case, the selection by memory is linearly preferential in space, since the walker chooses a visited position with a probability proportional to the total amount of time spent at that position.

The whole range of possible memory kernels was studied in \cite{BEM2017}. It was shown that in the case of strong memory of the initial condition, where $K(\tau,t)$ is strongly peaked at $\tau=0$,  the probability distribution of the walker's position reaches
a non-Gaussian stationary state, which contains memory of the initial condition and depends in a complicated way on the whole evolution of the walk.
On the other hand, when $K(\tau,t)$ decreases more slowly than $1/\tau$ or increases, the late time behaviour is a time-dependent distribution of Gaussian form, in which the initial condition is less and less revisited and where the variance $\sigma^2(t)$ of the position grows unbounded.
The variance may take a 
remarkable range of dynamical behaviours depending on the form of the memory kernel. The behaviours
range from ultra-slow growth of the variance  $\sigma^2(t) \sim \ln \ln t$ in the borderline case $K(\tau,t) \sim 1/\tau$, through
slow, logarithmic growth when $K(\tau,t) \sim \tau^\alpha$ with $\alpha >-1$, to diffusive behaviour with a modified diffusion constant when
$ K(\tau,t)  \sim \exp(a \tau)$ with $a>0$.
Thus, memory can fundamentally change the diffusive dynamics. Rigorous results using
the properties of weighted random recursive trees have confirmed some of these predictions
\cite{MUB2019,BMa2023,BMa2024}.

An interesting question is how these different dynamical behaviours are modified when a confining potential is introduced.
For a simple diffusive process in a confining potential, the long-time distribution 
approaches a Gibbs-Boltzmann stationary state
and the relaxation towards this state is typically an  exponential decay at late times, where the decay rate is determined by the potential and the diffusion coefficient of the particle.
Recently, for linear preferential relocation in  the case $K(\tau,t) =1/t$ 
(uniformly selected $\tau$)  it was shown that 
while the final stationary state is still Gibbs-Boltzmann,
the relaxation to the stationary state is very different: it is sluggish and decays
as a power law in time
with a non-trivial exponent that depends on the relocation rate $r$ as well as on 
the details of the potential \cite{BM2024}.

\begin{figure}[t]
\centering
\includegraphics[width=0.8\textwidth]{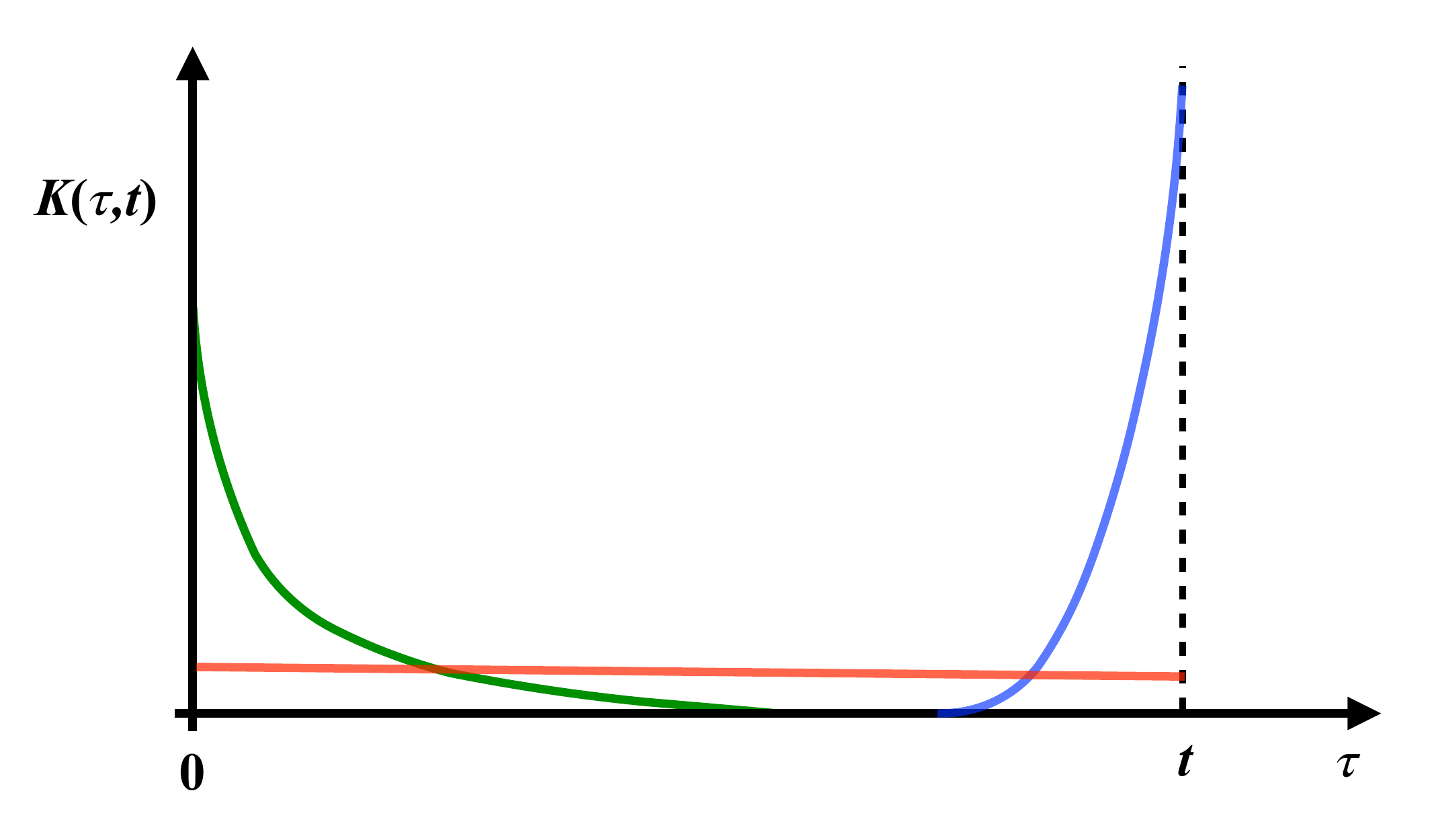}
\caption{Memory kernels. At time $t$, a confined diffusive particle chooses (with rate $r$) a time $\tau\in[0,t]$ with probability density $K(\tau,t)$, and revisits the position occupied at that time $\tau$. Fixing $t$, a rapidly decaying $K(\tau,t)$ with $\tau$ indicates that only the initial part of the trajectory is remembered (green curve), whereas with a kernel peaked near the present time the walker remembers recent positions better (blue curve). Another possibility is that all the times are equally sampled (red curve).  Taking $K(\tau,t)\propto \phi(\tau)$, two asymptotic dynamical regimes emerge: if $\phi(\tau)$ decays faster than $1/\tau$ (kernel localized near $\tau=0$) the stationary state is strongly affected by the initial position $x_0$ and the resetting rate $r$. If $\phi(\tau)$ decreases slower than $1/\tau$ or increases, the stationary state is the Gibbs-Boltzmann distribution, independent of $x_0$ or $r$. }
\label{fig:diagram}
\end{figure}

In this  work we consider the preferential relocation model with a general memory kernel in a confining potential. We present a general
theory for the long-time behaviour and confirm the predictions of the theory in some exactly solvable cases.
Our key findings are that for strong memory of the initial condition, the stationary state is non-Gaussian and depends on the initial condition. In this case we refer to the memory kernel $K(\tau,t)$ as being {\em localized} near $\tau=0$ as illustrated in Fig.~\ref{fig:diagram}.
We refer to the opposite case, when $K(\tau,t)$ decreases more slowly than $1/\tau$ or
increases, as a delocalized memory kernel---see Fig.~\ref{fig:diagram}. (Note that this also includes the cases where
$K(\tau,t)$ is peaked at $t$.)
For a delocalized memory kernel a variety of slow relaxation dynamics to a Gibbs-Boltzmann
stationary state are exhibited.
Our theory reveals that the general form of the relaxation behaviour is determined by the memory kernel in a rather simple way. The confining potential affects the details of the relaxation, e.g., when the relaxation is of a power law form, it enters into the expression of the exponent.

The paper is organised as follows. In Section~\ref{sec:model} we define the
preferential relocation model in a confining potential. In Section~\ref{sec:sep} we
use separation of variables to reduce the problem of determining the probability distribution into a space-dependent part and a time-dependent part.
In Section~\ref{sec:asymp} we present an asymptotic theory for the time-dependent part of the solution and categorise the possible behaviors. In Section~\ref{sec:exact} we exactly solve certain cases from which we verify the predictions of the asymptotic theory. Section \ref{sec:simul} displays simulation results in support of the theory. We conclude in Section~\ref{sec:conc}.

\section{Model Definition}\label{sec:model}

We consider a particle that starts at a position $x_0$ at time $t=0$ and diffuses in continuous time on a line in the presence of a 
confining potential $U(x)$. In addition to Brownian diffusion,
the particle undergoes a memory-dependent resetting with rate $r$. More precisely, at time $t$ the particle located at position $x(t)$ has the whole history of its own trajectory $\{x(\tau)\}_{0\le \tau\le t}$ 
at its disposal. The position $x(t)$ is updated in a small time $\Delta t$ according to the following
protocol. With probability $r\Delta t$ the particle decides to reset and with the complementary probability $1-r \Delta t$,
it diffuses. If it  resets, it chooses any previous time $0\le \tau\le t$ from its history with a probability density $K(\tau,t)$, where $\int_0^t K(\tau,t)d\tau=1$,
and resets to the position $x(\tau)$. The memory is not refreshed after relocation. The probability density $p(x,t)$ of the particle's position
then evolves via the Fokker-Planck equation
\begin{equation}
\fl
\frac{\partial p(x,t)}{\partial t}= D\, \frac{\partial^2 p(x,t)}{\partial x^2}+ 
\frac{\partial}{\partial x}\left[U'(x)\, p(x,t)\right] -r\, p(x,t) + r\, \int_0^t K(\tau,t)\, p(x,\tau)\, d\tau\, ,
\label{fp.1}
\end{equation}
starting from the initial condition $p(x,0)= \delta(x-x_0)$. The first two terms on the right hand side (rhs) describe
the standard diffusion of a particle with diffusion constant $D$ and friction coefficient set to unity in an external potential $U(x)$. If the friction coefficient is $\gamma$, Eq. (\ref{fp.1}) applies with the change $U(x)\to U(x)/\gamma$. The third term on the rhs corresponds to
the loss from the position at $x$ due to resetting (with rate $r$) to other positions. The last term corresponds to
the gain in the probability density at $x$ due to resetting to $x$ from other positions. To understand this term more clearly,
let $p(x',t; x, \tau)$ denote the joint probability density that the particle is at $x'$ at time $t$ and $x$ at time $\tau\le t$. Then, from this current position $x'$ at $t$, it can reset with rate $r$ to the position $x$
with probability $K(\tau,t) d\tau$. The total probability flux to the position $x$ due to resetting from all the possible 
positions $x'$ occupied at time $t$ is given by
\begin{equation}
r\, \int_0^t \left[ \int_{-\infty}^{\infty}  p(x',t; x,\tau) dx'\right] K(\tau,t)\, d\tau= r\, \int_0^t K(\tau,t)\, p(x,\tau)\, d\tau \,,
\label{flux.1}
\end{equation}
where we used the fact that $\int_{-\infty}^{\infty} p(x',t; x,\tau)dx'= p(x,\tau)$ is the marginalised
position distribution at time $\tau$. This explains the last term on the rhs of Eq. (\ref{fp.1}).
Using the condition $\int_0^t K(\tau,t)\, d\tau=1$ and the initial condition $p(x,0)=\delta(x-x_0)$, it is easy to check
that Eq. (\ref{fp.1}) conserves the total probability, i.e., $\int_{-\infty}^{\infty} p(x,t)\, dx=1$.

Here we will focus on a particular family of kernels $K(\tau,t)$ of the form \cite{BEM2017,MUB2019,BMa2024}
\begin{equation}
K(\tau,t)= \frac{\phi(\tau)}{I(t)}\, ,
\label{kernel.1}
\end{equation}
with
\begin{equation}
  I(t)= \int_0^t \phi(\tau)  d\tau\, ,\label{Idef}
\end{equation}
where $\phi(\tau)$ is any non-negative function of $\tau$. Thus the kernel $K(\tau,t)$ depends on $t$ only through
the denominator $I(t)$. For this choice,
the Fokker-Planck equation (\ref{fp.1}) reduces to
\begin{equation}
\fl
\frac{\partial p(x,t)}{\partial t}= D\, \frac{\partial^2 p(x,t)}{\partial x^2}+
\frac{\partial}{\partial x}\left[U'(x)\, p(x,t)\right] -r\, p(x,t) + \frac{r}{I(t)}\, 
\int_0^t \phi(\tau)\, p(x,\tau)\, d\tau %\, , \quad {\rm where}\quad I(t)=\int_0^t \phi(\tau)\, d\tau\, .
\label{fp.2}
\end{equation}
Note that, since ${\dot I}(t)=\phi(t)\ge 0$ is non-negative, $I(t)$ can never decrease with time $t$. There are thus only two possible late time
behaviours of $I(t)$: (I) $I(t)$ increases without limit as $t\to \infty$, and (II) 
$I(t)$ approaches a constant from below as $t\to \infty$.
This specific class of resetting memory kernel was introduced
in Ref.~\cite{BEM2017} where the position distribution $p(x,t)$ was analysed in detail, but in the absence of any external confining potential $U(x)$.

The function $\phi(\tau)$ includes two well-studied cases:
 $\phi(\tau)=1$ where all times $\tau$ prior to $t$ are sampled
uniformly for resetting~\cite{BS2014} and  $\phi(\tau)=\delta(\tau)$ which corresponds to the standard  Poissonian resetting~\cite{EM2011,EMS2020} which always occurs to the initial position $x_0$. The first case $\phi(\tau) =1$ is an example of a delocalized memory kernel and the second case  $\phi(\tau)=\delta(\tau)$ is an example of a localized kernel.
It was found in Ref.~\cite{BEM2017} that in the absence of $U(x)$, there are two different late time behaviours for $p(x,t)$
depending on the two cases (I) and (II) discussed above.

\begin{itemize}

\item {\bf Case (I): Delocalized memory kernel.} This case corresponds to when $\phi(\tau)$ decays as or slower than $1/\tau$ for large $\tau$, i.e., $I(t)=\int_0^t \phi(\tau)\, d\tau$ increases indefinitely with $t$ for large $t$. If $U(x)=0$, then $p(x,t)$ remains time dependent
even at long times (no stationary state) and the growth of the mean square displacement exhibits a variety of late time growth behaviours  (sub-diffusive and diffusive) depending on the details of $I(t)$ for large $t$ \cite{BEM2017}.
 
\item {\bf Case (II):  Localized memory kernel. } This complementary case corresponds 
to the one where $\phi(\tau)$ decays faster than $1/\tau$ for large $\tau$,
i.e., $\tau\, \phi(\tau)\to 0$ as $\tau\to \infty$. In this case $I(t)$ approaches a constant
from below as $t\to \infty$. If $U(x)=0$, then $p(x,t)$ approaches a stationary state as $t\to \infty$
that depends explicitly on the resetting rate $r$
\cite{BEM2017}.

\end{itemize}

The purpose of this paper is to study how these behaviours are affected when there is an additional external
confining potential $U(x)$. There are two special cases of $\phi(\tau)$ for which $p(x,t)$, in the
presence of a confining potential $U(x)$, has been studied.

\begin{enumerate}

\item $\phi(\tau)=1$. In this case, it was shown recently~\cite{BM2024} that $p(x,t)$ approaches
a stationary state at long times that has the Gibbs-Boltzmann form where detailed balance is satisfied,
\begin{equation}
p_{\rm st}(x) = \frac{1}{Z}\, e^{-U(x)/D}\, \quad {\rm where} \quad Z= \int_{-\infty}^{\infty} e^{-U(x)/D}\,dx\, ,
\label{Gibbs.1}
\end{equation}
provided $U(x)$ is sufficiently confining such that $Z$ is finite. The stationary state is thus independent of
the resetting rate $r$. Hence in this case the delocalized memory kernel fails to affect the long time stationary distribution of the particle when the 
confining $U(x)$ is switched on. However,
the effect of  memory shows up in an anomalous relaxation to the stationary state. It was shown that 
$p(x,t)-p_{\rm st}(x)\sim t^{-\theta_1}$
decays as a power law at late times with an exponent $\theta_1$ that
depends continuously on $r$ and the potential parameters.
%This should be contrasted with the....

\item $\phi(\tau)=\delta(\tau)$. This corresponds to diffusion with Poissonian 
resetting~\cite{EM2011, EMS2020} to the initial condition. In this case,  when the external potential $U(x)$ is switched on, it was shown~\cite{P2015}
that the system reaches a stationary state
at long times, but the stationary distribution $p_{\rm st}(x)$ has a non-Boltzmann form that depends explicitly
on the resetting rate $r$, as well as on $U(x)$.
\end{enumerate} 

In this paper, we generalise these results to the case of arbitrary confining potential $U(x)$ and arbitrary
non-negative $\phi(\tau)$. Our main results can be summarized as follows.
Our first observation is that Eq. (\ref{fp.2}) [and in fact more generally Eq. (\ref{fp.1})] always admits a time-independent `fixed-point' solution
\begin{equation}
p_{\rm Gibbs}(x)= \frac{1}{Z}\, e^{-U(x)/D}\, \quad {\rm where} \quad Z= \int_{-\infty}^{\infty} e^{- U(x)/D}\,dx\, ,
\label{Gibbs.2}
\end{equation}
assuming $Z$ is finite. This is easily seen by a direct substitution of Eq. (\ref{Gibbs.2}) into Eq. (\ref{fp.1}).
The first two terms on the rhs of Eq. (\ref{fp.1}) cancel each other, and so do the third and the last term upon using
the fact that $\int_0^{t} K(\tau,t)\, d\tau=1$. Now the interesting question is whether the actual
solution $p(x,t)$ approaches this Gibbs-Boltzmann fixed point  at long times or not. We show in this paper
that this depends on the tail of $\phi(\tau)$ or equivalently that of $I(t)=\int_0^t \phi(\tau)d\tau$.
More specifically, we show that there is always a stationary state at long times in the
presence of $U(x)$ for any non-negative $\phi(\tau)$. However, the nature of this stationary state is very 
different in the two cases (I) and (II) above: % Our main findings are summarized below.

\begin{itemize}

\item In case (I) where $I(t)$ increases indefinitely with $t$ at late times, i.e., 
$\phi(\tau)$ decays as or slower than $1/\tau$ for large $\tau$ (or increases), the stationary state is indeed given by
the Gibbs-Boltzmann form in Eq. (\ref{Gibbs.2}),
for sufficiently confining $U(x)$ such that $Z$ is finite. 
This thus generalises the result for $\phi(\tau)=1$ mentioned in  point (i) above.
This means that as long as $I(t)$ increases indefinitely with $t$, even
though the memory-induced resetting is delocalized in the absence of $U(x)$, it 
doesn't affect the Gibbs-Boltzmann stationary state at long times when $U(x)$ is 
switched on. Essentially, in this case,
at large $t$, the integral in the last term on the rhs of Eq. (\ref{fp.2}) is dominated by large $\tau$ where $p(x,\tau)$
is independent of $\tau$ and hence to leading order for large $t$ the last term in Eq. (\ref{fp.2}) 
behaves as $r\, p_{\rm st}(x)$ which then cancels the third term on the rhs of Eq. (\ref{fp.2}). Furthermore,
$p_{\rm st}(x)= p_{\rm Gibbs}(x)$ then makes the sum of the first two terms vanish, showing
that at late times $p(x,t)$ indeed converges to the Gibbs-Boltzmann state (\ref{Gibbs.2}). 
Thus the potential $U(x)$
overrides, in some sense, the memory effect due to resetting as far as the final stationary state
is concerned. 
The delocalized memory in resetting however
affects the {\em relaxation} to this stationary state and we find a variety of relaxation behaviours to the
Gibbs-Boltzmann state depending on how $I(t)$ increases with $t$ at late times. 
This is an example of weakly broken ergodicity where the system does indeed explore the full space to
ultimately reach the Gibbs-Boltzmann state, but the time to reach this stationary state can be very long depending
on how $I(t)$ grows with $t$ for large $t$. 
Interestingly, in this case, the system approaches a Gibbs-Boltzmann stationary distribution even
though the detailed balance is, in general, not satisfied.

\item In case (II) where $I(t)$ approaches a constant from below at late times, i.e., 
$\phi(\tau)$ decays faster than $1/\tau$ for large $\tau$, the stationary state
is non-trivial and has a non-Boltzmann form that depends on $r$, $U(x)$, $\phi(t)$ and also
on the initial condition. Thus in this case, even though the Gibbs-Boltzmann state (\ref{Gibbs.2}) is
a fixed point, it is an `unstable' fixed point. The system does not reach this Gibb-Boltzmann state at late times
and rather converges to a different non-trivial fixed point. This is thus an example of strongly
broken ergodicity, where the initial condition is remembered at all times, all the way to the stationary state.

\end{itemize}

\section{Solution via the method of separation of variables}
\label{sec:sep}

In the absence of the potential $U(x)$, the system is homogeneous in space and the
Fokker-Planck equation (\ref{fp.2}) can be solved for arbitrary $\phi(\tau)$
in the Fourier space~\cite{BEM2017}. However a non-zero $U(x)$ makes the space
inhomogeneous and the method using Fourier transform is no longer appropriate.
Instead, an alternative method using separation of variables
is more suitable to solve Eq. (\ref{fp.2}). In fact, this method was already used
in Ref.~\cite{BM2024} to solve the special case $\phi(\tau)=1$, i.e., $I(t)=t$.
Below we show that the same method can be used for general $\phi(\tau)$.

To solve the Fokker-Planck equation (\ref{fp.2}) via separation of variables we first
seek a particular product solution of the form
\begin{equation}
p(x,t)= g_{\lambda}(x)\, f_{\lambda}(t)
\label{sep_var.1}
\end{equation}
where $\lambda$ is a constant (independent of space and time). Substituting Eq. (\ref{sep_var.1}) in (\ref{fp.2}), dividing both sides by $g_{\lambda}(x)\, f_{\lambda}(t)$, and assembling the space-dependent and the time-dependent parts separately gives
\begin{equation}
\fl
\frac{D\, g_\lambda''(x) + \partial_x\left[U'(x)\, g_\lambda(x)\right]}{g_\lambda(x)}=
\frac{\dot f_\lambda(t)}{f_\lambda(t)}+ r - \frac{r}{f_\lambda(t)\, I(t)}\, \int_0^t 
\phi(\tau)\, f_\lambda(\tau)\, d\tau=-\lambda\, .
\label{sep_var.2}
\end{equation}

\vskip 0.3cm

\noindent {\bf The space-dependent part.} Let us first consider the space-dependent part $g_\lambda(x)$ that satisfies
the second order eigenvalue equation
\begin{equation}
D\, g_\lambda''(x)+ \partial_x\left[U'(x)\, g_\lambda(x)\right]= -\lambda\, g_\lambda(x), 
\label{space.1}
\end{equation}
where $\lambda\ge 0$ plays the role of an eigenvalue yet to be determined. The restriction
$\lambda\ge 0$ will automatically emerge from a mapping to a Schr\"odinger problem
as follows.
One can reduce this eigenvalue equation to a familiar Schr\"odinger form by making
the substitution~\cite{Risken}
\begin{equation}
g_\lambda(x) = e^{-\frac{1}{2D} U(x)}\, \psi_\lambda(x)\, .
\label{space.2}
\end{equation}
It is then easy to verify that $\psi_\lambda(x)$ satisfies the standard Schr\"odinger equation (with $m=\hbar=1$)
\begin{equation}
-\frac{1}{2} \psi_\lambda''(x)+ V_Q(x)\, \psi_\lambda(x)= \frac{\lambda}{2D}\, \psi_\lambda(x)\, ,
\label{space.3}
\end{equation}
where the quantum potential $V_Q(x)$ is given by~\cite{Risken}
\begin{equation}
V_Q(x)= -\frac{1}{4D} U''(x)+ \frac{1}{8 D^2}\, \left(U'(x)\right)^2\, .
\label{quantum_pot.1}
\end{equation} 
Thus to compute the space-dependent part $g_\lambda(x)$ one needs to solve the Schr\"odinger equation (\ref{space.3}),
find the spectrum $\lambda_n\ge 0$, i.e., the eigenvalues $E_n= \lambda_n/(2D)\ge 0$ and the associated eigenfunctions
$\psi_{\lambda_n}(x)$. Note that the Schr\"odinger problem does not allow negative eigenvalues $E_n$'s.
Moreover, the quantum potential $V_Q(x)$ in Eq. (\ref{quantum_pot.1}) is such that its ground state occurs
with eigenvalue $E_0=\lambda_0/(2D)=0$ and the corresponding normalized eigenfunction is 
simply~\cite{Risken}
\begin{equation}
\psi_0(x)= a_0 e^{-\frac{1}{2D} U(x)}\, ,\quad {\rm with}\quad a_0=\int_{-\infty}^{\infty}dx\ e^{-\frac{1}{D} U(x)}
\label{ground.1}
\end{equation}
as can be seen by setting $\lambda =0$ in Eq. (\ref{space.1}).
Note also that, in general, the spectrum of the Schr\"odinger equation with $V_Q(x)$ may consist of both
bound and scattering states, even though the classical potential $U(x)$ may be fully confining~\cite{Risken}. 
Finally, we note that this space-dependent part is completely independent of 
resetting since $r$ and $\phi(\tau)$ does not appear
in the space-dependent solution $g_\lambda(x)$. The resetting only affects the time-dependent part
$f_\lambda(t)$ which we consider next.

\vskip 0.3cm

\noindent {\bf The time-dependent part.} For a given $\lambda$ (which is fixed by the spatial part
of the solution, i.e., by the diffusion problem in the absence of resetting), the time-dependent part $f_\lambda(t)$ of the solution satisfies the first order integro-differential equation
\begin{equation}
{\dot f}_\lambda(t) + (r+\lambda)\, f_\lambda(t)= \frac{r}{I(t)}\, \int_0^{t} \phi(\tau)\, f_\lambda(\tau)d\tau\, ,
\label{time.1}
\end{equation}
starting from some initial condition $f_\lambda(0)$. Actually, this equation can be reduced
to an ordinary second order differential equation by taking one more derivative of Eq. (\ref{time.1}) with respect to $t$
and using the relation ${\dot I}(t)=\phi(t)$. This gives
\begin{equation}
{\ddot f}_\lambda(t) +\left[r+\lambda+ \frac{{\dot I}(t)}{I(t)}\right]\, {\dot f}_\lambda(t)+ \lambda \, 
\frac{{\dot I}(t)}{I(t)}\, f_\lambda(t)=0\, ,
\label{time.2}
\end{equation}
satisfying the initial conditions
\begin{equation}
f_\lambda(t=0)= f_\lambda(0)\, , \quad {\rm and}\quad {\dot f}_\lambda(t=0)=-\lambda\, f_\lambda(0)\, .
\label{init_cond}
\end{equation}
The last condition follows by taking the $t\to 0$ limit of the ratio of integrals in the rhs of  Eq. (\ref{time.1}).

For a general $I(t)$ (or equivalently $\phi(t)$), the second order differential
equation (\ref{time.2}) is hard to solve exactly for all $t$. We will see later that some special choices of $I(t)$
allow for the exact solution of Eq. (\ref{time.2}). However, we will show that the behaviour of $f_\lambda(t)$ for large
$t$ depends on the form of $I(t)$ at large $t$ and can be inferred for arbitrary $I(t)$.
Let us point out one important fact. For the ground state $\lambda=0$, the time-dependent part $f_0(t)$ satisfies
\begin{equation}
{\ddot f}_0(t) +\left[r+ \frac{{\dot I}(t)}{I(t)}\right] {\dot f}_0(t)=0\, ,
\label{f0t.1}
\end{equation}
which can be solved for arbitrary $I(t)$. The general solution reads
\begin{equation}
f_0(t)= A - B\, \int_{t}^{\infty} dt' \frac{ e^{-r t'}}{I(t')}\, .
\label{f0t_sol.1}
\end{equation}
where $A$, $B$ are constants.
The constant $B$ is fixed to be zero by the initial condition $\dot f_\lambda(t=0) =0$, thus
%Since $I(t)$ can either increase with time indefinitely
%or at most approach a constant as $t\to \infty$, it follows that as $t\to \infty$,
%the function $f_0(t)$ approaches a constant as $t\to \infty$
\begin{equation}
f_0(t)=A\, .
\label{f0t_final.1}
\end{equation}
Hence, the Gibbs-Boltzmann form (\ref{ground.1}) associated to $\lambda=0$ is also stationary in the presence of memory.

Having obtained both the spatial and the temporal part of the solution, we can then write down the general
solution of the Fokker-Planck equation (\ref{fp.2}) as a linear combination over all allowed eigenvalues
\begin{equation}
p(x,t) =\sum_{n} a_{\lambda_n}\, \left[ e^{-\frac{1}{2D}\, U(x)} \psi_{\lambda_n}(x)\right]\, f_{\lambda_n}(t)\, ,
\label{gen_sol.1}
\end{equation}
where the unknown coefficients $a_{\lambda_n}$ are fixed by the initial condition $p(x,t=0)$. Thus, given
an external potential $U(x)$, one needs
to first solve the Schr\"odinger equation (\ref{space.3}) and obtain its full spectrum of eigenvalues
$\lambda_n$'s and eigenfunctions $\psi_{\lambda_n}(x)$'s. Next, for a given $\lambda_{n}$, one needs
to solve the differential equation (\ref{time.2}) to obtain the temporal part $f_{\lambda_n}(t)$.
Finally, one needs to fix the constants $a_n$ in Eq. (\ref{gen_sol.1}) from the initial condition by using the orthogonality property $\int_{-\infty}^{\infty}dx\ \psi_{\lambda_n}(x)\psi_{\lambda_m}(x)=\delta_{n,m}$.
These three steps then provide the full exact solution (\ref{gen_sol.1}). 
In the next section, we will provide a general theory of relaxation of $p(x,t)$ at late times based on the 
large time behaviour 
of $f_{\lambda_n}(t)$ and this theory is valid for arbitrary $I(t)$. 
Later in the paper, we will provide some exactly solvable examples of $I(t)$ for which
$f_{\lambda_n}(t)$ can be obtained at all times. These exactly solvable examples will
provide supporting evidence for the general late time solution developed in the next section.

\section{Asymptotic solution at late times}\label{sec:asymp}

For a generic $I(t)$, we have seen above from Eq. (\ref{f0t_final.1}) that $f_0$ is a constant. Consequently, one can separate out the ground state $\lambda=0$ in
Eq. (\ref{gen_sol.1}) from the rest of the spectrum and re-write it as
\begin{equation}
p(x,t) = a_0 e^{-\frac{1}{D}\, U(x)} + \sum_{\lambda_n>0} a_{\lambda_n}\,
\left[ e^{-\frac{1}{2D}\, U(x)} \psi_{\lambda_n}(x)\right]\, f_{\lambda_n}(t)\, ,
\label{gen_sol.2}
\end{equation}
where we used the fact that $\psi_0(x)= e^{-U(x)/(2D)}$ from Eq. (\ref{ground.1}). Note that
the first term in Eq. (\ref{gen_sol.2}) corresponds exactly to the Gibbs-Boltzmann stationary state and indeed it is
a fixed point solution (as $t\to \infty$) of Eq. (\ref{fp.2}) irrespective of $I(t)$. Now, whether the
system approaches this fixed point at late times depends on how the rest of the sum in Eq. (\ref{gen_sol.2})
behaves at late times. We will see below that in case (I),  where $I(t)$ increases with time as $t\to \infty$, 
the contribution from the higher excited states [i.e., the sum with $\lambda_n>0$ in Eq.  (\ref{gen_sol.2})]
vanishes as $t\to \infty$, ensuring that the system indeed converges to the Gibbs-Boltzmann stationary state (\ref{Gibbs.2})
at late times. In contrast, in case II where $I(t)$ approaches a constant as $t\to \infty$, the sum over $\lambda_n>0$
in Eq. (\ref{gen_sol.2}) remains non-zero as $t\to \infty$, unless the initial distribution $p(x,t=0)$ is exactly $a_0 e^{-U(x)/D}$ itself. Consequently, the system generically approaches a
non-trivial stationary state that is characterized by the full spectrum of the associated Schr\"odinger
operator and not just by the ground state as in case (I). 

\subsection{Case (I): $I(t)$ increases with $t$ as $t\to \infty$}

We start with case (I), where $I(t)$ increases with time at late times. We need to analyse how the
sum over the contributions from the excited states ($\lambda_n>0$) in Eq. (\ref{gen_sol.2})
behaves as $t\to \infty$. For this, we need to understand how $f_{\lambda_n}(t)$ in Eq. (\ref{time.2})
with $\lambda_n>0$ behaves at late times. In this case, anticipating that $f_{\lambda_n}(t)$
decreases to zero as $t\to \infty$ for $\lambda_n>0$ , we can neglect the second derivative ${\ddot f}_{\lambda_n}(t)$
at late times. This, of course, needs to be justified {\it a posteriori}. The approximate evolution
equation at late times becomes
\begin{equation}
\left[r+\lambda+ \frac{{\dot I}(t)}{I(t)}\right] {\dot f}_\lambda(t)+ \lambda \,
\frac{{\dot I}(t)}{I(t)}\, f_\lambda(t)\approx 0\, .
\label{ftI.1}
\end{equation}
Now, there are two subcases: (Ia)  $\frac{{\dot I}(t)}{I(t)}$ decreases to zero with $t$ or (Ib)
$\frac{{\dot I}(t)}{I(t)}$ increases to infinity as $t\to \infty$. Below we consider the two cases 
separately.

\vskip 0.3cm

\noindent {\bf The case (Ia).} In this case, while $I(t)$ increases with $t$, the ratio
$\frac{{\dot I}(t)}{I(t)}$ decreases with increasing $t$ and vanishes as $t\to \infty$.
At late times, we can further neglect the ratio $\frac{{\dot I}(t)}{I(t)}$ inside
the parenthesis compared to the constant term $(r+\lambda)$, leading to
\begin{equation}
(r+\lambda) {\dot f}_\lambda(t)\approx - \lambda \,
\frac{{\dot I}(t)}{I(t)}\, f_\lambda(t)\, .
\label{casea.1}
\end{equation}
Solving this equation, we find that at late times
\begin{equation}
f_{\lambda_n}(t) \approx \frac{b_n}{\left[I(t)\right]^{ \frac{\lambda_n}{r+\lambda_n} }}, 
\label{casea.2}
\end{equation}
where $b_n$ is a constant.  To verify {\it a posteriori} that 
neglecting ${\ddot f}_{\lambda_n}(t)$ is justified for this solution,
one  obtains from Eq. (\ref{casea.1})
\begin{equation}
 \frac{{\ddot f}_\lambda(t)}{f_\lambda(t)}\approx \left[ \left(\frac{\lambda}{r+\lambda}\right)^2\
\left(\frac{{\dot I}(t)}{I(t)}\right)^2  -  \left(\frac{\lambda}{r+\lambda}\right)\
\frac{d}{dt} \left( \frac{{\dot I}(t)}{I(t)}\right)\right], \,\label{casea.3}
\end{equation}
and one sees that the rhs vanishes as $t\to\infty$.
Consequently, for all $I(t)$'s belonging to this subcase (Ia),  the late time dependence 
$f_{\lambda_n}(t)$ of each excited mode $\lambda_n>0$ vanishes as 
\begin{equation}
f_{\lambda_n}(t) \sim \left[I(t)\right]^{-\frac{\lambda_n}{r+\lambda_n}}\, .
\label{f_I_asymp.1}
\end{equation}
Therefore
the sum over the excited states in Eq. (\ref{gen_sol.2}) also vanishes, and the system approaches
the Gibbs-Boltzmann state (\ref{Gibbs.2}) asymptotically. However, the relaxation to this Gibbs-Boltzmann state depends
on the precise form of the memory kernel $I(t)$ and decays as a power of $I(t)$, $[I(t)]^{-\lambda_n/(r+\lambda_n)}$,
at late times. We now consider some examples of case (Ia).

Our first example is the choice
\begin{equation}
I(t) \approx b\, t^{\alpha}\, , \quad {\rm with}\quad \alpha>0\, ,
\label{ex1.1}
\end{equation}
where $b>0$ is a constant. In this case the ratio decreases as $\frac{{\dot I}(t)}{I(t)}\approx 1/t$
as $t\to \infty$ and it is easy to see that Eq. (\ref{casea.3}) vanishes as $t\to \infty$, thus the
asymptotic behaviour (\ref{f_I_asymp.1}) holds.
We find that from Eq. (\ref{casea.2}) that for all $\lambda_n>0$, the time-dependent contribution from
this excited mode slowly decays at late times as a power law
\begin{equation}
  f_{\lambda_n}(t) \sim t^{-\alpha \lambda_n/(r+\lambda_n)}\;. \label{casea.ex1}
\end{equation}
Note that this case includes the special case $\alpha=1$ or $I(t)=t$, i.e., $\phi(\tau)=1$, of the linear preferential relocation model \cite{BM2024}.  If the eigenvalues $\lambda_n$ are discrete, as is usual, and ordered as $0<\lambda_1<\lambda_2<\ldots$,
the function $f_{\lambda_n}(t)$ with the smallest non-zero exponent will dominate the relaxation in the sum (\ref{gen_sol.1}) at large $t$.
We deduce that the late time relaxation to the steady state is anomalous and dominated by the first excited mode $n=1$,
\begin{equation}
p(x,t)-p(x,t\to \infty)\sim t^{-\alpha\lambda_1/(r+\lambda_1)},
\label{rel.pow}
\end{equation}
if the initial condition is such that $\psi_{\lambda_1}(x_0)\neq 0$.
The result of \cite{BM2024} is recovered by setting $\alpha=1$ in Eq. (\ref{rel.pow}). 
In fact  $I(t) = b\, t^\alpha$ with $\alpha>0$ is an exactly solvable case that we
will present in Section~\ref{sec:exact}, showing that the exact solution
reproduces the late time behaviour predicted by our asymptotic theory.

Another example of case (Ia) is $I(t) \sim e^{\gamma t^\beta}$
with $0<\beta<1$. In this example $\dot I/I \simeq \gamma \beta
t^{\beta-1}$ and one can check that Eq. (\ref{casea.3}) vanishes as $t\to
\infty$.
The  time-dependent contribution from
this excited mode decays at late times as
\begin{equation}
 f_{\lambda_n}(t) \sim e^{-\gamma
   \lambda_n/(r+\lambda_n)\,t^\beta}
 \end{equation}
i.e. a stretched exponential decay. Similarly, for a discrete spectrum the asymptotic decay is controlled by the mode $n=1$, or
\begin{equation}
p(x,t)-p(x,t\to \infty)\sim e^{-\gamma\lambda_1/(r+\lambda_1)\,t^\beta}.
\label{rel.str}
\end{equation}

\vskip 0.3cm

\noindent {\bf The case (Ib).} In this case, both $I(t)$ and the ratio $\frac{{\dot I}(t)}{I(t)}$ increase with
time for large $t$. An example of this case is the choice $I(t)\sim  e^{\gamma\, t^{\beta}}$ with $\beta>1$.
Then the ratio behaves as $\frac{{\dot I}(t)}{I(t)}\approx \gamma \beta t^{\beta-1}$ which grows with time
for all $\beta>1$. 
Consequently, one can neglect the constant term $(r+\lambda_n)$ inside
the parenthesis in the first term in Eq. (\ref{ftI.1}), leading to a late time solution
\begin{equation}
f_{\lambda_n}(t)\approx c_n\, e^{-\lambda_n\, t}\, .
\label{caseb.1}
\end{equation}
Thus for all $I(t)$'s belonging to this class, the time-dependent contribution from an excited mode $\lambda_n>0$
decays exponentially as $e^{-\lambda_n\, t}$, which is the same behaviour as that of the memory-less process (with $r=0$) and does not depend on the details of $I(t)$. Hence, resetting
in this case does not affect even the relaxation rates (the amplitudes may still depend on resetting). In other words, the memory kernel is most ineffective and the system behaves as if it was just the normal diffusion
in the confining potential $U(x)$. Given that all the excited modes vanish as $t\to \infty$,
the solution $p(x,t)$ in Eq. (\ref{gen_sol.2}) again converges to the Gibbs-Boltzmann stationary 
state (\ref{Gibbs.2}), and $p(x,t)-p(x,t\to\infty)\sim e^{-\lambda_1 t}$ for a discrete spectrum. It happens that the case with $\beta=2$, or $I(t)=e^{{\gamma}t^2}$, is exactly solvable and its analysis will be presented in Section \ref{sec:exact}, too.\\

\noindent{\bf Borderline case}. There is also an interesting borderline case between (Ia) and (Ib) when $I(t)$ increases
exponentially at late times, $I(t) \sim e^{\gamma t}$ with $\gamma >0$. In this case, the ratio
$\frac{{\dot I}(t)}{I(t)}\approx \gamma$ approaches a constant and one can no longer neglect the second derivative term ${\ddot f}_\lambda(t)$
in Eq. (\ref{time.2}). At late times, using $\frac{{\dot I}(t)}{I(t)}\approx \gamma$, Eq. (\ref{time.2}) reduces to
\begin{equation}
{\ddot f}_\lambda(t) +[r+\lambda+ \gamma]\, {\dot f}_\lambda(t)+ \lambda \,
\gamma\, f_\lambda(t)=0\, .
\label{time_border.1}
\end{equation}
This equation can be trivially solved giving
\begin{equation}
f_\lambda(t)= A_+\, e^{-\kappa_+ t} + A_{-}\, e^{-\kappa_{-} t}\, ,
\label{ft_border_sol.1}
\end{equation}
where $A_{\pm}$ are arbitrary constants and
\begin{equation}
\kappa_{\pm}= \frac{1}{2}\left[r+\lambda+\gamma\pm \sqrt{(r+\lambda+\gamma)^2- 4 \lambda \gamma}\right]\, .
\label{kappa_def}
\end{equation}
%Since $\kappa_{-}<0$ and the solution can not diverge exponentially as $t\to \infty$, we must choose
%$A_{-}=0$.
As both $\kappa_\pm$ are real and positive, the solution at late times for an excited mode $\lambda_n>0$ decays as
\begin{equation}
f_{\lambda_n}(t)\sim e^{-\kappa_-t}\, .
\label{ft_border_sol.2}
\end{equation}
Thus in this borderline case, the solution $p(x,t)$ again converges to the Gibbs-Boltzmann state
at late times, and the relaxation to this stationary state is exponential as in case (Ib) above.
However, unlike in (Ib), the relaxation rate $\kappa_-$ {\em does} depend explicitly on the resetting rate $r$.

\subsection{Case (II): $I(t)$ approaches a constant from below as $t\to \infty$}

In this case, $\phi(t)={\dot I}(t)$ decreases with time. Consequently,  
the ratio $\frac{{\dot I}(t)}{I(t)}$ will decrease with time and can be neglected compared
to the constant $(r+\lambda)$ in the parenthesis of the second term in Eq. (\ref{time.2}).
This leads to an approximate equation at late times
\begin{equation}
{\ddot f}_\lambda(t) +\left[r+\lambda\right]\, {\dot f}_\lambda(t)
\approx - \lambda\, \frac{ {\dot I}(t)}{I(t)}\, f_\lambda(t)\, .
\label{ft_II.1}
\end{equation}
To make further progress, we anticipate (and verify {\it a posteriori}) that $f_\lambda(t)$ approaches
a constant $f_\lambda(\infty)$ as $t\to \infty$. Assuming this, we can replace $f_\lambda(t)$ at late times
by its asymptotic value $f_\lambda(\infty)$ on the rhs of Eq. (\ref{ft_II.1}). Defining
${\dot f}_\lambda(t)= v_\lambda(t)$, we obtain a first order inhomogeneous equation for $v_\lambda(t)$
which can be solved explicitly giving 
\begin{equation}
v_\lambda(t) = A_1\, e^{-(r+\lambda)\, t} - \lambda f_\lambda(\infty)\, e^{-(r+\lambda)\, t}\,
\int_{t_0}^{t}  \frac{{\dot I}(t')}{I(t')}\ e^{(r+\lambda)\, t'}\, dt'\, ,
\label{vlt.1}
\end{equation}
where $A_1$ is a constant and $t_0\sim O(1)$ is a microscopic time scale beyond which
the asymptotic solution is expected to be valid.
Furthermore, since $\frac{{\dot I}(t')}{I(t')}$ is a decreasing function of $t'$, we can integrate
the second term by parts and to leading order at late times we get
\begin{equation}
{\dot f}_\lambda(t)= v_\lambda(t) \approx A_1\, e^{-(r+\lambda)\, t} - 
\frac{\lambda\, f_\lambda(\infty)}{(r+\lambda)}\, \frac{{\dot I}(t)}{I(t)}\, .
\label{vlt.2}
\end{equation}
Integrating further with respect to $t$ gives
\begin{equation}
f_\lambda(t) \approx f_\lambda(\infty)+ \frac{A_1}{(r+\lambda)}\, e^{-(r+\lambda)\, t}
+  \frac{\lambda\, f_\lambda(\infty)}{(r+\lambda)}\, \ln\left[\frac{I(t)}{I(\infty)}\right]\, .
\label{ft_II.2}
\end{equation}
Since $I(t)\to I(\infty)$ as $t\to \infty$, this proves self-consistently that indeed $f_\lambda(t)$
approaches a constant $f_\lambda(\infty)$ as $t\to \infty$ for all $\lambda\ge 0$.
However, the nature of the temporal decay to this final constant value depends on the nature of $I(t)$.
If $\ln[I(t)/I(\infty)]$ decays faster than $e^{-(r+\lambda)\, t}$ then the leading relaxation is
controlled by $e^{-(r+\lambda)\, t}$ and in the opposite case it is controlled by $\ln[I(t)/I(\infty)]$.
In the next section, we will provide an exact solution for the case $\phi(t)= \mu\, e^{-\mu t}$
which will demonstrate this point clearly.

Thus, $f_\lambda(t)$ approaches a constant at late times for any $\lambda\ge 0$.
Consequently, from Eq. (\ref{gen_sol.2}), we see that as $t\to \infty$, all the modes contribute
to $p(x, t\to \infty)$. Hence, the system reaches a new stationary state
\begin{equation}
p_{\rm st}(x)= \sum_{n} a_{\lambda_n}\, \left[ e^{-\frac{1}{2D}\, U(x)} 
\psi_{\lambda_n}(x)\right]\, f_{\lambda_n}(\infty)\, ,
\label{st_II.1}
\end{equation}
which is no longer the simple Gibbs-Boltzmann state (\ref{Gibbs.2}). This is a nonequilibrium stationary state
(since the resetting violates detailed balance in general) and this stationary state
depends explicitly on the potential $U(x)$, the resetting rate $r$, the localized memory kernel $I(t)$ and
on the initial condition.
In the next section, we will compute $p(x,t)$ at all times for two solvable examples
belonging to this class (II) of $I(t)$ that will indeed support this general conclusion.

\section{Exactly solvable examples of $\phi(t)$}\label{sec:exact}

In this section, we provide a few examples of $\phi(t)$, or equivalently of $I(t)=\int_0^t \phi(\tau) d\tau$,
for which $f_\lambda(t)$ in Eq. (\ref{time.2}) can be solved explicitly at all times $t$.
These solvable cases will include examples belonging to both classes (I) and (II).

\subsection{The case $I(t)=b\, t^{\alpha}$}
As a first example belonging to the case (Ia) in the previous section, we consider for all $t\ge 0$
\begin{equation}
I(t)= b\, t^{\alpha}\, \quad {\rm with}\quad \alpha>0 \, .
\label{ex1_exact.1}
\end{equation}
For this choice, the ratio ${\dot I}(t)/I(t)= \alpha/t$ for all $t$. Consequently, Eq. (\ref{time.2}) reduces to
\begin{equation}
t\, {\ddot f}_{\lambda}(t) + \left[\alpha+ (r+\lambda)t\right] {\dot f}_{\lambda}+ \alpha\, \lambda\, f_\lambda(t)=0\, ,
\label{ex1.2}
\end{equation}
valid for all $t\ge 0$ and subject to the initial conditions in Eq. (\ref{init_cond}). This equation is of the form of Kummer's differential equation \cite{AS_book} \begin{equation} 
z W''(z) + (b-z)\, W'(z)- a\, W(z)=0\, , \label{Kummer_ODE} 
\end{equation}
with $z=-(r+\lambda)t$, $a=\alpha\lambda/(r+\lambda)$ and $b=\alpha$.  The two linearly independent solutions of Eq. (\ref{Kummer_ODE}) are Kummer's confluent hypergeometric functions $M(a,b,z)$ and $U(a,b,z)$, where the first one is defined by the series expansion  
\begin{equation}
M(a,b,z)= 1+ \frac{a}{b} z+ \frac{a(a+1)}{b(b+1)}\, \frac{z^2}{2!}+ \frac{a(a+1)(a+2)}{b(b+1)(b+2)}\frac{z^3}{3!}+\ldots\, .
\label{Kummer_def.1}
\end{equation} and $U(a,b,z)$ is its
linearly independent cousin \cite{AS_book}.  The general solution of Eq. (\ref{ex1.2}) is thus given by the linear combination
\begin{equation}
f_\lambda(t)= C_1\, M\left(\frac{\alpha\lambda}{r+\lambda}, \alpha, - (r+\lambda)t\right)
+ C_2\, U\left(\frac{\alpha\lambda}{r+\lambda}, \alpha, - (r+\lambda)t\right)\, ,
\label{ex1.3}
\end{equation}
The constants $C_1$ and $C_2$ are fixed by the two
initial conditions in Eq. (\ref{init_cond}).
In fact, since $dU(a,b,z)/dz \sim z^{-b}$ as $z\to 0$, we must
have $C_2=0$, leading to the exact solution at all times
\begin{equation}
f_\lambda(t)= f_\lambda(0)\, M\left(\frac{\alpha\lambda}{r+\lambda}, \alpha, - (r+\lambda)t\right)\, .
\label{ex1.4}
\end{equation}
It is easy to check that Eq. (\ref{ex1.4}) satisfies the initial
conditions (\ref{init_cond}) by using the definition (\ref{Kummer_def.1}).
Now, as $t\to \infty$, using the asymptotic behaviour
\begin{equation}
M(a,b,z)\approx \frac{\Gamma(b)}{\Gamma(b-a)}\, (-z)^{-a}\, , \quad {\rm as}\quad z\to -\infty \, ,
\label{M_asymp.1}
\end{equation}
 we find that to leading order for large $t$ 
\begin{equation}
f_\lambda(t)\approx f_\lambda(0)\, \frac{\Gamma(\alpha)}{ \Gamma\left(\frac{\alpha\, r}{r+\lambda}\right) }\, 
\left[(r+\lambda)\, t\right]^{- \frac{\alpha \lambda}{r+\lambda}}\, .
\label{ex1.5}
\end{equation}
Since $I(t)$ is given by Eq. (\ref{ex1_exact.1}), we can rewrite the expression (\ref{ex1.5}) as
$f_\lambda(t)\sim \left[I(t)\right]^{-\lambda/(r+\lambda)}$
as $t\to \infty$, in perfect agreement with the general prediction in Eq. (\ref{f_I_asymp.1}).

\subsection{The case $I(t)=e^{\gamma t^2}$}

This case $I(t)= e^{\gamma t^2}$ for all $t$ 
provides an example for the class (Ib) discussed in the previous section. In this case, Eq. (\ref{time.2})
reduces to
\begin{equation}
{\ddot f}_\lambda(t) +\left[r+\lambda+ 2\,\gamma\, t\right]\, {\dot f}_\lambda(t)+ 2\lambda \,
\gamma\, t\, f_\lambda(t)=0\, ,
\label{ex2.1}
\end{equation}
subject to the two initial conditions in Eq. (\ref{init_cond}).
To solve the differential equation (\ref{ex2.1}),
%we need to first reduce it to some known form via some transformation
it turns out to be useful first to make the substitution \begin{equation} f_\lambda(t)= 
e^{-r t- \gamma\, t^2}\, F_\lambda(t)\, . \label{ex2.2} \end{equation} Then $F_\lambda(t)$ 
satisfies \begin{equation} {\ddot F}_\lambda(t) +\left[-r +\lambda -2 \gamma\, 
t\right]\,{\dot F}_\lambda(t)- (2\gamma+r \lambda)\, F_\lambda(t)=0 \, . \label{ex2.3} 
\end{equation} Furthermore, let us make a change of variable (after suitable inspection) 
\begin{equation} z= \left( \frac{r-\lambda}{2\sqrt{\gamma}}+ \sqrt{\gamma}\, t\right)^2\, . 
\label{ex2.4} \end{equation} Then it is easy to check that $F_\lambda(t)= W(z)$ satisfies 
the differential equation \begin{equation} z\, W''(z) + \left(\frac{1}{2}-z\right)\, W'(z) - 
\left(\frac{1}{2}+ \frac{r\lambda}{4\gamma}\right)\, W(z)=0\, . \label{ex2.5} \end{equation} 
This is once again of the form of Kummer's differential equation (\ref{Kummer_ODE}), whose two linearly independent solutions are $M(a,b,z)$ and $U(a,b,z)$ defined previously. Hence, 
putting together all these results, we have the exact solution of Eq. (\ref{ex2.1}) given by 
\begin{eqnarray} 
f_{\lambda}(t)= e^{-rt-\gamma\, t^2}&&\left[C_1\, M\left(\frac{1}{2}+ 
\frac{r\lambda}{4\gamma}, \frac{1}{2},
 \left( \frac{r-\lambda}{2\sqrt{\gamma}}+ \sqrt{\gamma}\, t\right)^2\right)\right.\nonumber\\
&&\left. +C_2\, U\left(\frac{1}{2}+ \frac{r\lambda}{4\gamma}, \frac{1}{2},
\left( \frac{r-\lambda}{2\sqrt{\gamma}}+ \sqrt{\gamma}\, t\right)^2\right)\right]\, ,
\label{ex2.6}
\end{eqnarray}
where the unknown constants $C_1$ and $C_2$ are fixed by the two initial conditions
 (\ref{init_cond}). To derive the asymptotic late time behaviour of
$f_\lambda(t)$ in  Eq. (\ref{ex2.6}), we use 
the following asymptotic behaviours for large $z$~\cite{AS_book} 
\begin{equation}
M(a,b,z) \approx \frac{\Gamma(b)}{\Gamma(a)}\, z^{a-b}\, e^{z}\, \quad {\rm and}\quad
U(a,b,z)\approx z^{-a}\, .
\label{MU_largez}
\end{equation}
This gives the leading large $t$ asymptotic behaviour of $f_\lambda(t)$
\begin{equation}
f_\lambda(t) \approx \left[\frac{C_1\, \sqrt{\pi} }{\Gamma\left(\frac{1}{2}+\frac{r\lambda}{4\gamma}\right)}\,
\, \gamma^{\frac{r\lambda}{4\gamma}}\, e^{(r-\lambda)^2/(4\gamma)}\right]\, 
t^{\frac{r\lambda}{2\gamma}}\, e^{-\lambda t}\, .
\label{ex2.7}
\end{equation}
This leading exponential decay (up to pre-exponential power law growth) is in perfect agreement with the general prediction in Eq.
(\ref{caseb.1}).  

\subsection{The case $I(t)= 1- e^{-\mu\,t}$}\label{sec:expon}

In this case we choose $\phi(t)= \mu\, e^{-\mu t}$ or $I(t)= 
\int_0^t \phi(\tau)\, d\tau= 1-e^{-\mu\, t}$.
Since $I(t)\to 1$ from below as $t\to \infty$, this is an example belonging to the class (II) in
the previous section. In this case, Eq. (\ref{time.2}) reduces to
\begin{equation}
{\ddot f}_\lambda(t) + \left[r+ \lambda + \frac{\mu\, e^{-\mu t}}{1-e^{-\mu t}}\right]\, {\dot f}_\lambda(t)
+ \lambda\,  \frac{\mu\, e^{-\mu t}}{1-e^{-\mu t}}\, f_\lambda(t)=0\, ,
\label{ex3.1}
\end{equation}
subject again to the initial conditions (\ref{init_cond}).
To solve this equation, we again need to first reduce to a known form using appropriate change of variables.
Indeed, defining $y= e^{-\mu t}$, one finds that $f_\lambda(t)=G(y)$ satisfies the equation
\begin{equation}
y(1-y)\, G''(y) + \left[1- \frac{r+\lambda}{\mu}- 
\left(2-\frac{r+\lambda}{\mu}\right)\,y\right]\,
G'(y) + \frac{\lambda}{\mu}\, G(y)=0\, .
\label{Gy.1}
\end{equation}
This form is the familiar hypergeometric differential equation
\begin{equation}
y(1-y)\, G''(y) + \left[c- (a+b+1)\, y\right]\, G'(y) - ab\, G(y)=0\, ,
\label{Gy.2}
\end{equation}
with the identification of the three parameters
\begin{equation}
\fl
a= \frac{1}{2}\left[c+\sqrt{c^2+\frac{4\lambda}{\mu}}\right]\, , \quad\, 
b= \frac{1}{2}\left[c-\sqrt{c^2+\frac{4\lambda}{\mu}}\right] \quad\,
{\rm and}\quad c= 1- \frac{r+\lambda}{\mu}\, .
\label{def_abc}
\end{equation}
The general solution of Eq. (\ref{Gy.2}) then can be expressed as a linear combination
\begin{equation}
G(y)= B_1\, F(a,b,c,y) + B_2\, y^{1-c}\, F(a-c+1, b-c+1, 2-c, y)\, ,
\label{Gy_sol.1}
\end{equation}
where $B_1$ and $B_2$ are arbitrary constants and $F(a,b,c,z)$ is the hypergeometric series
\begin{equation}
\fl
F(a,b,c,z)= 1+ \frac{ab}{c}\, z+ \frac{a(a+1)\, b(b+1)}{c(c+1)}\, \frac{z^2}{2!}+ 
\frac{a(a+1)(a+2)\, b(b+1)(b+2)}{c(c+1)(c+2)}\, \frac{z^3}{3!}+\ldots\, .
\label{hyper_def}
\end{equation}
Therefore, the exact solution of Eq. (\ref{ex3.1}), valid at all times $t$, is given by
\begin{equation}
f_\lambda(t)= B_1\, F\left(a,b,c, e^{-\mu t}\right)+ B_2\, e^{-(r+\lambda)\, t}\, F\left(a-c+1, b-c+1, 2-c, e^{-\mu t}\right)\, ,
\label{ex3.2}
\end{equation}
where the constants $B_1$ and $B_2$ are fixed from the initial conditions (\ref{init_cond})
and $a$, $b$, $c$ are given in Eq. (\ref{def_abc}). It can be verified \cite{BEM2017}  that the
two constants $B_1$ and $B_2$ are given by
\begin{equation}
B_1= f_\lambda(0)\, \frac{\Gamma(1-c)}{\Gamma(1-a)\Gamma(1-b)}\, , \quad {\rm and}
\quad B_2= - f_\lambda(0)\, \frac{\Gamma(c)}{(1-c)\, \Gamma(a)\Gamma(b)}\, .
\label{B1B2}
\end{equation}
In the long time limit $t\to \infty$, using $y=e^{-\mu t}\to 0$ and the series expansion (\ref{hyper_def}),
one finds that (keeping the leading and the next subleading terms)
\begin{eqnarray}
f_\lambda(t) \approx&& f_\lambda(0)\, \frac{\Gamma(1-c)}{\Gamma(1-a)\Gamma(1-b)}
+ f_\lambda(0)\, \frac{a\, b\,\Gamma(1-c)}{c\, \Gamma(1-a)\Gamma(1-b)}\, e^{-\mu t}\nonumber\\
&&- f_\lambda(0)\, \frac{\Gamma(c)}{(1-c)\, \Gamma(a)\Gamma(b)}\, e^{-(\lambda+r)\, t} \, .
\label{ex3.3}
\end{eqnarray}

The above exact result indeed demonstrates that at long times $f_\lambda(t)$ approaches
a constant given by the first term in Eq. (\ref{ex3.3}) whose value depends on the initial condition
$f_\lambda(0)$ and also on the parameters $r$ and $\mu$. Moreover, the leading asymptotic decay
to this constant is either $e^{-(r+\lambda)\, t}$ as given by the third term, or
by $e^{-\mu t}$ (the second term) depending on whether $(r+\lambda)$ is smaller or larger than $\mu$.
This result is in perfect agreement with the general result presented for case (II)
in Eq. (\ref{ft_II.2}). Indeed, using $I(t)=1- e^{-\mu t}$, we see that 
$\ln\left[\frac{I(t)}{I(\infty)}\right]\approx -e^{-\mu t}$ at late times. 
%Hence theexact result in (\ref{ex3.3}) corroborates the general result derived in (\ref{ft_II.2}).
 
\subsection{The case $\phi(t)=\delta(t)$ corresponding to standard resetting}\label{sec:reset}

We now consider the case of standard resetting where resetting always occurs to the initial 
position $x=0$. This case corresponds to the choice $\phi(\tau)=\delta(\tau)$, or equivalently
$I(t)=1$. Thus this example also belongs to  class (II) for localized memory kernels in the previous section.
In fact, this is a special case of $\phi(t)=\mu\, e^{-\mu t}$ in the limit $\mu\to \infty$.
But it is instructive to study this case separately, as the first order integro-differential equation (\ref{time.1}) simplifies to
\begin{equation}
{\dot f}_\lambda(t) + (r+\lambda)\, f_\lambda(t)= r\, f_\lambda(0)\, ,
\label{reset.1}
\end{equation}
which can be explicitly solved to give, at all times, 
\begin{equation}
f_\lambda(t)= f_\lambda(0)\, \left[\frac{r}{r+\lambda}+ 
\frac{\lambda}{r+\lambda}\, e^{-(r+\lambda)\, t}\right]\, .
\label{reset.2}
\end{equation}
Consequently, the exact solution for $p(x,t)$ in Eq. (\ref{gen_sol.1}) reads
\begin{equation}
p(x,t)= \sum_{\lambda_n} \frac{ f_{\lambda_n}(0)}{(r+\lambda_n)}\, \left[ r+ \lambda_n\, 
e^{-(r+\lambda_n)\, t}\right]\,
\left[ e^{-\frac{1}{2D}\, U(x)} \psi_{\lambda_n}(x)\right]\, ,
\label{reset.3}
\end{equation}
where we have absorbed the constants $a_{\lambda_n}$ in $f_{\lambda_n}(0)$ without any loss
of generality. From Eq. (\ref{reset.3}), it is clear that as $t\to \infty$, all the modes
contribute leading to a non-trivial nonequilibrium stationary state
\begin{equation}
p_{\rm st}(x)= p(x, t\to \infty)= \sum_{\lambda_n} \frac{r\, f_{\lambda_n}(0)}{(r+\lambda_n)}
\left[ e^{-\frac{1}{2D}\, U(x)} \psi_{\lambda_n}(x)\right]\, ,
\label{reset.4}
\end{equation}
again in agreement with the general result derived in Eq. (\ref{st_II.1}).

In fact, in this standard resetting case the probability density $p(x,t)$ can
also be derived by an alternative renewal approach. In the appendix we show how the exact solution in Eq. (\ref{reset.3}), obtained via the method of separation of variables,
is consistent with the renewal method.

\section{Numerical Simulations}\label{sec:simul}

\begin{figure}[t]
\centering
\includegraphics[width=0.95\textwidth]{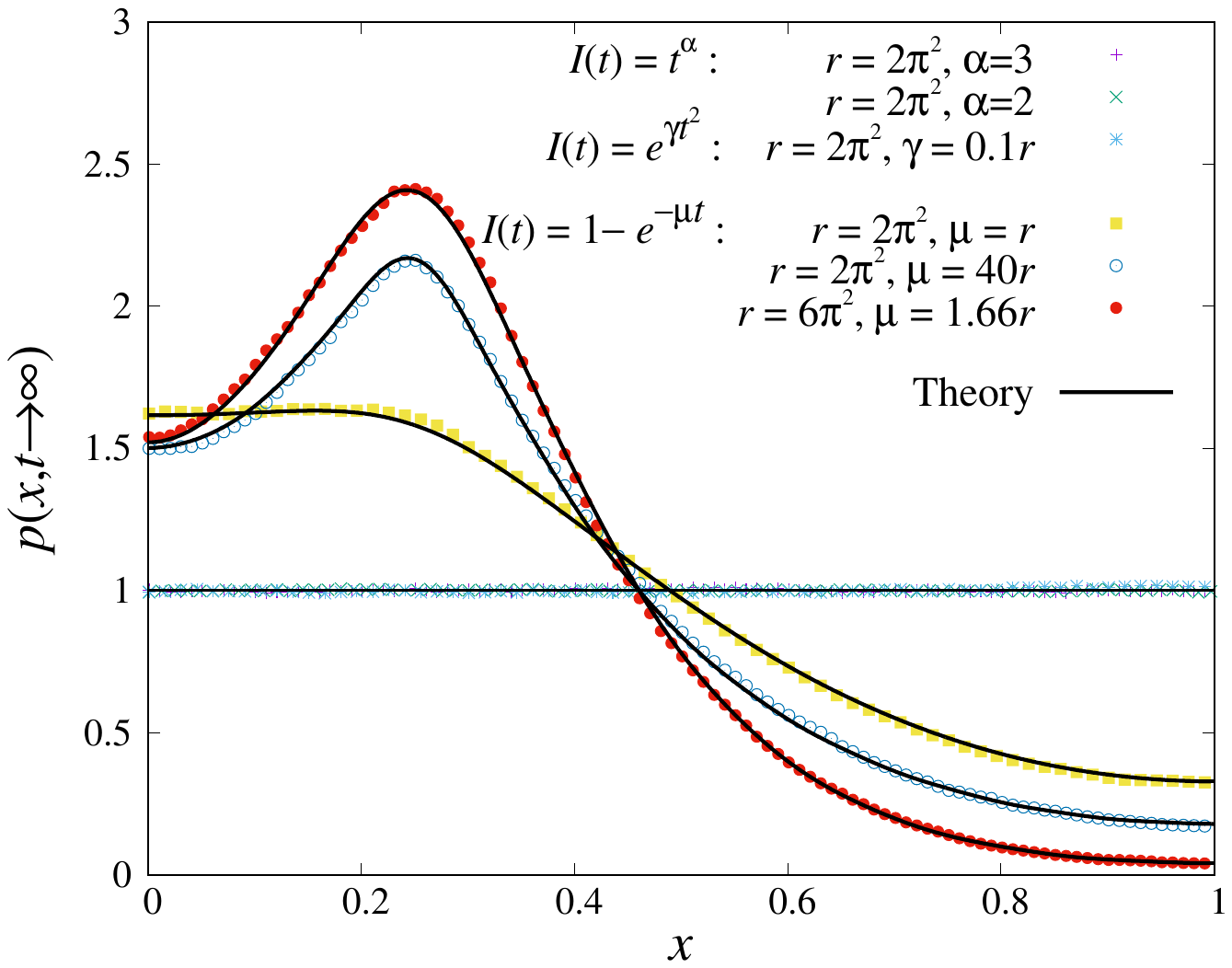}
\caption{Simulation results: asymptotic position density of a particle confined in a box potential $U(x)=0$ for $0< x < L$ with  reflecting boundary conditions at  $x=0$ and $x=L$.  Results for  different memory kernels indicated in the legend are presented. The parameters are $L=1$, $D=1$, $x_0=0.25$, the simulation time-step is $\Delta t=10^{-6}$ and the final time is set to $100$ or $400$. The averages are performed over the last 10 time units of $10^3$ independent trajectories. Depending on the behaviour of $I(t)$ at large time, $p(x, t\to \infty)$ tends toward the uniform Gibbs-Boltzmann distribution (crosses) or a non-trivial steady state (circles). The solid lines represent the exact expression (\ref{steadyboxexp}).}
\label{fig:box}
\end{figure}

To verify some of the predictions of the previous sections, we have run Brownian dynamics simulations in an external potential subject to the additional protocol of resetting with memory (see, e.g., \cite{BM2024} for details). The random times $\tau$ in the past are obtained by drawing random numbers $z$ uniformly distributed in $[0,1]$. From the identity $dz=K(\tau,t)d\tau$ and integrating, one obtains $z=[I(\tau)-I(0)]/I(t)$ or $I(\tau)=zI(t)+I(0)$. Inverting this relation in the different examples studied above gives $\tau$ as a function of $z$. For the power law case $I(t)=bt^{\alpha}$, 
\begin{equation}
\tau=z^{1/\alpha}t,
\end{equation}
while for the exponential function $I(t)=1-e^{-\mu t}$, we have
\begin{equation}
\tau=-\frac{1}{\mu}\ln\left[1-z(1-e^{-\mu t})\right].
\end{equation}
The case $I(t)=e^{\gamma t^2}$ is a bit peculiar since $I(t=0)>0$, meaning that $\phi(\tau)$ has a singular part at $\tau=0$ and can be decomposed into $\phi(\tau)=I(0)\delta(\tau)+\phi_{reg}(\tau)$. Hence with a finite probability $I(0)/I(t)=e^{-\gamma t^2}$ the resetting event occurs exactly to the starting position $x_0$. With the complementary probability, resetting occurs to $x(\tau)$, where $\tau>0$ is now chosen by inverting the relation $I(\tau)=zI(t)$, which gives
\begin{equation}
\tau=\sqrt{t^2+\frac{1}{\gamma}\ln z},
\end{equation}
for $z\in[e^{-\gamma t^2},1]$. It can be noted that the probability to reset to $x_0$ rapidly becomes negligible as $t$ grows and most of the resetting times approach the present time $t$. 

\subsection{Steady state distributions}\label{simulsteady}

{\bf Box potential.} Our first example is a box potential where $U(x)=0$ for $0\le x\le L$ and 
$U(x)=\infty$ otherwise, corresponding to a particle trapped in an interval of length $L$ 
with reflective boundary conditions. As shown by Fig. \ref{fig:box}, in the examples 
$I(t)=bt^{\alpha}$ (with $\alpha=2$, $3$) and $I(t)=e^{\gamma t^2}$, the density at late 
times is given by the uniform Gibbs-Boltzmann distribution, which is the expected behaviour 
when $I(t)$ grows unbounded [classes (Ia) and (Ib) of Section \ref{sec:asymp}, 
respectively]. On the other hand, for the example $I(t)=1-e^{-\mu t}<1$ belonging to class 
(II), the density clearly reaches a non-trivial steady state that depends on $x_0$ and the 
parameters $r$ and $\mu$.

The steady state  in the case $I(t)=1-e^{-\mu t}$ can be obtained analytically by using the results of Section~\ref{sec:expon}. For the box potential, the $n$-th eigenmode (with $n=0,1,2,\ldots$) is 
given by $\psi_{\lambda_n}(x)=\cos(k_n x)$, where $k_n=n\pi/L$ ensures the no-flux condition 
at $x=0$ and $x=L$, while the $n$-th eigenvalue is $\lambda_n=D(n\pi/L)^2$. Imposing the 
initial condition to Eq. (\ref{gen_sol.1}) gives
\begin{equation}
\sum_{n=0}^{\infty}f_{\lambda_n}(0)\cos(k_nx)=\delta(x-x_0),
\end{equation}
where we have absorbed the constants $a_{\lambda_n}$ into $f_{\lambda_n}(0)$. From the orthogonality of the cosine functions, one gets $f_{0}(0)=1/L$ and $f_{\lambda_n>0}(0)=2\cos(k_nx_0)/L$. Inserting these expressions into the solution (\ref{ex3.3}) and taking $t\to\infty$ yields 
 $f_{\lambda_n}(\infty)$. Substituting the latter coefficients into Eq. (\ref{gen_sol.1}) gives the stationary state,
\begin{equation}\label{nessbox}
p(x,t\to \infty)=\frac{1}{L}+\frac{2}{L}\sum_{n=1}^{\infty}
\frac{\Gamma(1-c_n)\cos(k_nx_0)}{\Gamma(1-a_n)\Gamma(1-b_n)}\cos(k_n x),
\label{steadyboxexp}
\end{equation}
where $a_n$, $b_n$ and $c_n$ are given by the expressions 
in Eq. (\ref{def_abc}) with $\lambda$ replaced by $\lambda_n=D(n\pi/L)^2$. In the first term of the rhs of Eq. (\ref{steadyboxexp}), we have used the fact that $\Gamma(1-c_0)/[\Gamma(1-a_0)\Gamma(1-b_0)]=1$. The agreement between the exact stationary solution, Eq. (\ref{steadyboxexp}), and the Brownian dynamics simulations is very good in Fig. \ref{fig:box}. In the figure, the stationary probability distribution  is peaked around
the initial  position $x_0 = 0.25$. We recall that $\phi(\tau)= \mu \exp(-\mu \tau)$
and in the limit $\mu\to \infty$ one recovers the standard resetting case $\phi(\tau) \to \delta(\tau)$. Thus, in the large $\mu$ limit the stationary distribution has a cusp at $x_0$ \cite{EM2011}.
However, for finite $\mu$ the cusp is replaced by a smooth peak as seen in Fig.~\ref{fig:box}.

\begin{figure}[t]
\centering
\includegraphics[width=0.95\textwidth]{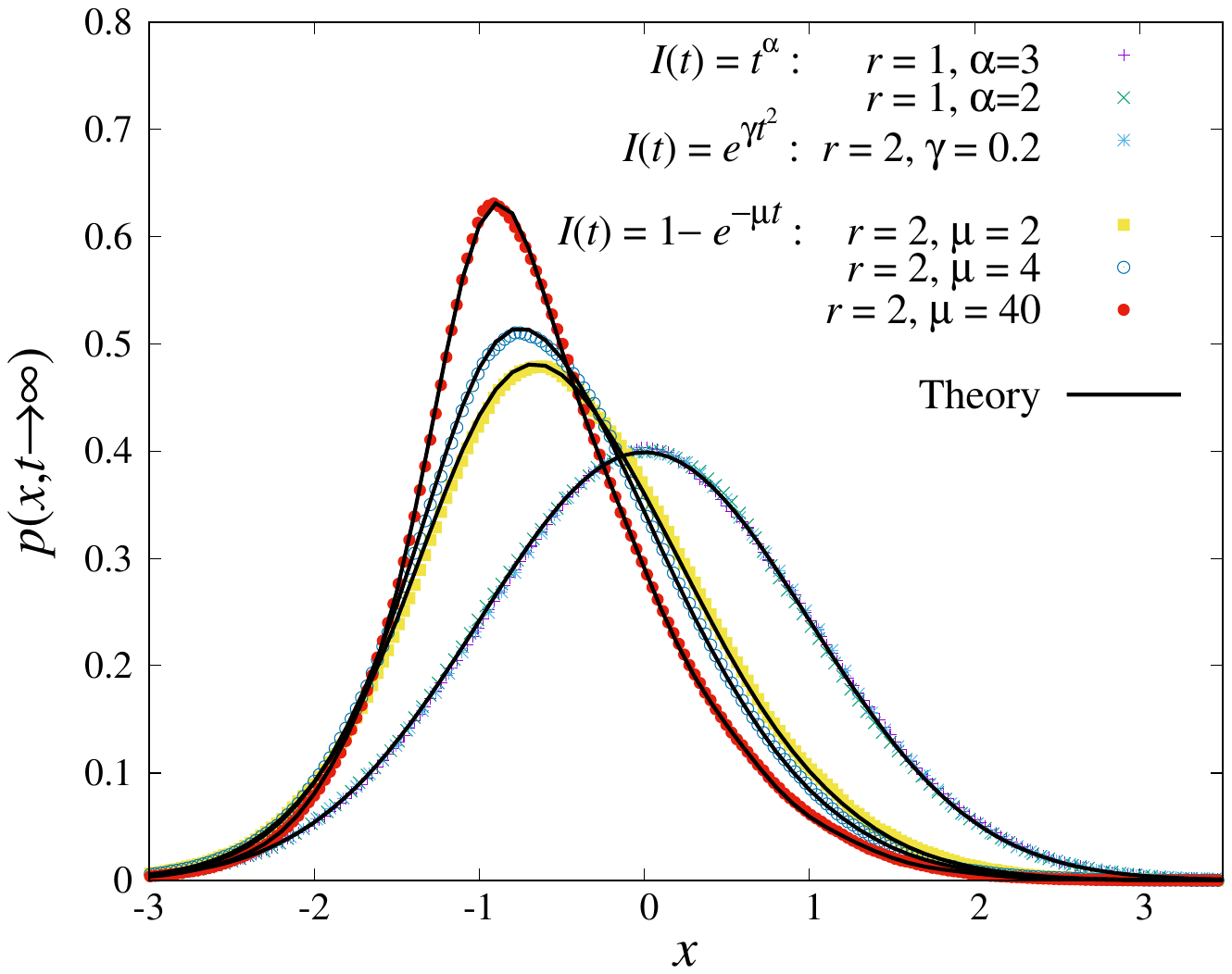}
\caption{Same as Figure \ref{fig:box} in the case of a particle with $D=1$ and initial position $x_0=-1$, in a harmonic potential $U(x)=\frac{1}{2}kx^2$ with $k=1$. Depending on the behaviour of $I(t)$
at large time, $p(x, t \to \infty)$ tends toward the Gaussian Gibbs-Boltzmann distribution
(crosses) or a non-trivial steady state (circles). The solid lines in the cases $I(t)=1-e^{-\mu t}$ represent the exact expression (\ref{nessOU}).}
\label{fig:OU}
\end{figure}

{\bf Harmonic potential.} Our second example is the harmonic potential $U(x)=\frac{1}{2}kx^2$, with $k$ a positive stiffness. For various memory kernels belonging to the cases (Ia) or 
(Ib), we verify in Figure \ref{fig:OU} that the density obtained numerically at late times (crosses) follows the Boltzmann-Gibbs distribution $\sqrt{\frac{k}{2\pi D}}e^{-\frac{k}{2D}x^2}$, which is indicated as a solid line.

With the kernel $I(t)=1-e^{-\mu t}$ of case (II), however, the density converges to a non-trivial distribution, as in the box potential. For any confining potential, one can use the orthogonality of the eigenfunctions $\psi_{\lambda_n}(x)$ to generalize the stationary density in Eq. (\ref{nessbox}) to
\begin{equation}\label{nessOU}
p(x,t\to \infty)=e^{\frac{1}{2D}[U(x_0)-U(x)]}\sum_{n=0}^{\infty} \frac{\Gamma(1-c_n)\psi_{\lambda_n}(x_0)}{\Gamma(1-a_n)\Gamma(1-b_n)}\psi_{\lambda_n}(x).
\end{equation}
In the case of the harmonic potential, the quantum potential (\ref{quantum_pot.1}) is also harmonic but shifted, i.e., $V_Q(x)=\frac{1}{2}\omega_0^2 x^2-\frac{1}{2}\omega_0$, with  $\omega_0=\frac{k}{2D}$. The eigenvalues of Eq. (\ref{space.3}) are thus $E_n=\omega_0(n+\frac{1}{2})-\frac{1}{2}\omega_0=\omega_0n$. Since $E_n=\frac{\lambda_n}{2D}$, one obtains
\begin{equation}\label{eigenvOU}
\lambda_n=kn,\quad {\rm with}\ n=0,1,2,\ldots
\end{equation}
The corresponding normalized eigenfunctions are given by \cite{CTDL1977},
\begin{equation}\label{eigenfOU}
\psi_{\lambda_n}(x)=\frac{1}{\sqrt{2^n n!}}\left( \frac{\omega_0}{\pi}\right)^{1/4} e^{-\frac{1}{2}\omega_0 x^2} H_n(\sqrt{\omega_0}x),
\end{equation}
where $H_n(z)$ are the Hermite polynomials \cite{AS_book}. The expression in Eq. (\ref{nessOU}) is evaluated numerically with Mathematica by combining Eqs. (\ref{def_abc}),  (\ref{eigenvOU}) and (\ref{eigenfOU}), and summing up to $n=80$. It is displayed in Fig. \ref{fig:OU} for different values of $\mu$ and an initial condition of $x_0=-1$ in reduced units (solid line). Once again, the theory agrees very well with the simulations.

\subsection{Relaxation toward the Gibbs-Boltzmann distribution}\label{simulrelax}

\begin{figure}[t]
\centering
\includegraphics[width=0.9\textwidth]{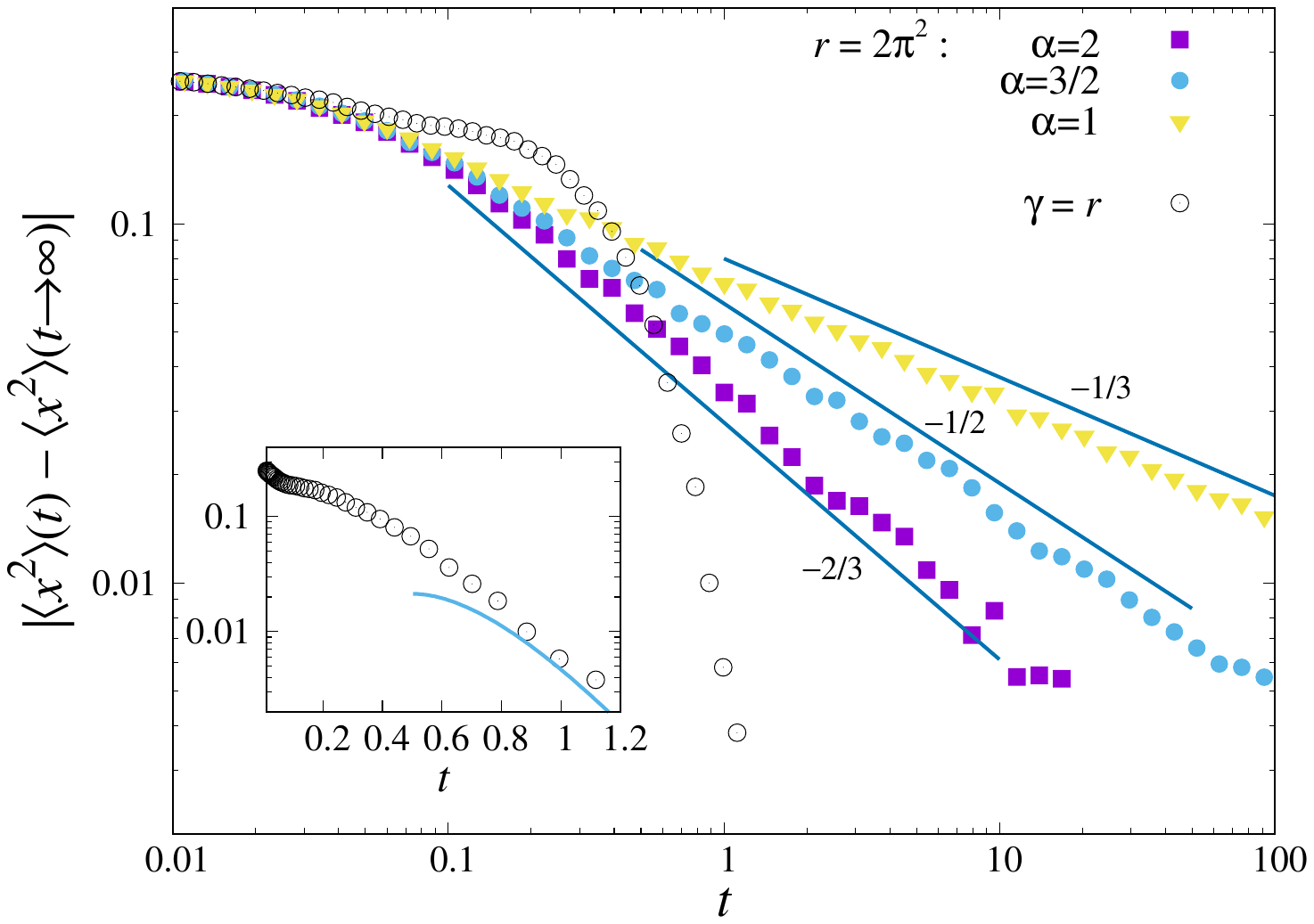}
\caption{Relaxation of the second moment of the position toward its asymptotic value (corresponding to the uniform distribution) for a particle confined by a box potential, with memory kernels $I(t)=b t^{\alpha}$ (filled symbols) and $I(t)= e^{\gamma t^2}$ (open circles). Averages are preformed over $10^4$ trajectories. Straight lines are guides to the eye and have the slope $-\alpha/3$ predicted from Eq. (\ref{rel.pow}). Unless indicated, the parameters are those of Figure \ref{fig:box}. Inset: semi-log plot of the curve corresponding to $I(t)= e^{\gamma t^2}$, where the solid line is given by Eq. (\ref{ex2.7}).}
\label{fig:boxrel}
\end{figure}

We now examine the relaxation dynamics toward the Boltzmann-Gibbs steady state, which differ in the cases (Ia) and (Ib). For convenience, we have only considered the box potential defined in Section \ref{simulsteady}. To avoid computing the full distribution $p(x,t)$, we have numerically obtained the second moment $\langle x^2\rangle(t)$, have assumed that $\langle x^2\rangle(t\to\infty)=L^2/3$ from the uniform distribution, and formed the quantity $|\langle x^2\rangle(t)-\langle x^2\rangle(t\to\infty)|$. This quantity is expected to have the same temporal behaviour as $p(x,t)-p(x,t\to\infty)$ at large time.

For the kernel $I(t)=bt^{\alpha}$, the leading behaviour of $p(x,t)-p(x,t\to\infty)$ is given by the inverse power law (\ref{rel.pow}) or (\ref{ex1.5}) where $\lambda_1=D(\pi/L)^2$ for a box. In Figure \ref{fig:boxrel}, we have set the resetting rate to $r=2\lambda_1$ and varied $\alpha$. As expected, the simulated relaxation is found to be algebraic and in very good agreement with the theoretical prediction $t^{-\alpha/3}$.

For the kernel of class (Ib) given by $I(t)=e^{\gamma t^2}$, the simulations of Fig. \ref{fig:boxrel} display a much faster relaxation (open circles and inset). The asymptotic exponential behaviour with the power law prefactor predicted in Eq. (\ref{ex2.7}) is close to the simulation results at large $t$.

\section{Summary and Conclusion}\label{sec:conc}

In this work we have considered the preferential relocation model
with a family of memory kernels given by Eq. (\ref{kernel.1}) in a general confining potential $U(x)$. Using
separation of variables we have reduced the problem of obtaining the
time-dependent probability distribution into a time-dependent part and
a space-dependent part. The space-dependent part reduces to solving a
Schr\"odinger equation with  quantum potential (\ref{quantum_pot.1}) to
determine the eigenvalues $\lambda_n \geq 0$ and corresponding
eigenfunctions $\psi_{\lambda_n}$.
The time-dependent part of the problem reduces to solving the second
order ordinary differential equation (\ref{time.2}) with initial
conditions (\ref{init_cond}). We have developed an asymptotic theory
for the solutions of this equation which allows us to predict and categorise the
late-time asymptotic behaviours of the probability distribution.

The asymptotic theory  identifies two generic  cases: (I) a delocalized memory kernel, when $\phi(\tau)$
decays as, or more slowly than $1/\tau$, for which  the probability distribution relaxes
to the  Gibbs-Boltzmann distribution; (II)  a localized memory kernel, when $\phi(\tau)$
decays more quickly than $1/\tau$, for which  the probability distribution
relaxes to a non-Gibbs-Boltzmann distribution containing memory of the
initial condition. 
The first case  (I) divides into two sub-cases: (Ia) in
which the relaxation is slower than exponential, such as a power law
or stretched exponential and (Ib) in which the
relaxation is exponential but with decay constant that depends on the
potential. It is interesting to note that in the absence of a potential these two cases correspond to the behaviour already
established of (i) a Gaussian distribution with variance $\sigma^2(t)$
depending on the delocalized memory kernel and (ii) a non Gaussian stationary
state with memory of the initial condition and dependence on the localized memory kernel.
Thus, a rich variety of relaxation behaviours are possible but the details of the relaxation
depend on the details of the memory kernel and confining potential.
The predicted behaviours are verified in a number of  cases where the
time-dependent part (\ref{time.2}) can be solved exactly, and by numerical simulations.

It would be of interest to extend these results to resetting in higher spatial dimensions
in the presence of a confining potential \cite{EM14}.
It would be also interesting to consider other memory kernels, for instance of the form $K(\tau,t)\propto \phi(t-\tau)$, that can be used to model how living organisms forget about the past \cite{BR2014}. The effects of all these memory kernels on the localization transitions of random walks on heterogeneous lattices deserve further study as well \cite{FCBGM2017}.

\section*{Acknowledgements}
We thank the hospitality of the Higgs Centre for Theoretical Physics at the University of 
Edinburgh during the workshop ``New Vistas on Stochastic Resetting" 
where this work was initiated. 
SNM acknowledges support from ANR Grant No. ANR-23-CE30-0020-01 EDIPS. DB acknowledges support from Conhacyt (Mexico) Grant CF2019/10872.

\appendix

\section{Relation between renewal approach and separation of variables} For the  case $\phi(\tau) = \delta(\tau)$ we demonstrate that
the separation of variables approach discussed in Section~\ref{sec:reset}  is equivalent
the result of a renewal equation approach commonly used n the literature.
Let us first recall briefly the
derivation of $p(x,t)$ using the renewal method~\cite{EMS2020}. Let $p_0(x,t)$
denote the probability density for a free diffusing particle (without resetting) in an external potential $U(x)$
to reach a point $x$ at time $t$, starting from the initial position $x=0$.
The subscript $0$ in $p_0(x,t)$ denotes $r=0$ (no resetting). Now, let us switch on the resetting with rate $r$.
This means that the intervals between successive resettings are distributed exponentially
with rate $r$~\cite{EM2011}, i.e., $\rho(\tau)= r\, e^{-r\, \tau}$ where $\tau$ denotes
the interval between two consecutive resettings. The successive resetting intervals are
statistically independent and hence the process renews itself everytime it resets. 
It is convenient to write down a renewal equation 
in terms of the last resetting interval before $t$~\cite{EMS2020}
\begin{equation}
p(x,t)= p_0(x,t)\, e^{-r\, t}+ r \int_0^{t} d\tau_l\, e^{-r\tau_l}\, p_0(x, \tau_l)\, ,
\label{renew.1}
\end{equation}
where the first term corresponds to the case with no resetting and the second term corresponds to one
or more resettings. In the latter case, let $\tau_l$ denote the interval between the last resetting and 
the current time $t$. If the particle has to reach $x$ at time $t$, it must propagate freely
during this last interval. This explains the factor $p_0(x,\tau_l)$ inside the integral. 
The probability that the interval $\tau_l$ contains no resetting event is simply $e^{-r\tau_l}$
and the probability that there is a resetting event inside the small interval $\left[t-\tau_l-d\tau_l, t-\tau_l\right]$
is simply $r\, d\tau_l$. Multiplying these probabilities and integrating over $\tau_l$ from $0$ to $t$
gives the second term. The convolution structure in the second term in Eq. (\ref{renew.1}) suggests
taking a Laplace transform with respect to $t$. We define
\begin{equation}
{\tilde p}(x,s)= \int_0^{\infty} p(x,t)\, e^{-s\, t}\, dt\, .
\label{laplace_def}
\end{equation}
Taking a Laplace transform of Eq. (\ref{renew.1}) gives
\begin{equation}
\tilde{p}(x,s) = \frac{(r+s)}{s}\, \tilde{p}_0(x, r+s)\, .
\label{laplace.1}
\end{equation}
   
Now, we ask if the solution (\ref{reset.3}) obtained by the method of separation 
of variables is consistent with the renewal solution (\ref{laplace.1}).
To check this, we first take the Laplace transform of Eq. (\ref{reset.3}). This gives
\begin{equation}
\tilde{p}(x,s)= \sum_{\lambda_n} \frac{f_{\lambda_n}(0)}{r+\lambda_n}\left[ \frac{r}{s}+
\frac{\lambda_n}{r+\lambda_n+s}\right]\, \left[ e^{-\frac{1}{2D}\, U(x)} \psi_{\lambda_n}(x)\right]\, .
\label{sepvar.1}
\end{equation}
Now we notice the identity
\begin{equation}
\frac{r}{s}+ \frac{\lambda_n}{r+\lambda_n+s}=\frac{(r+\lambda_n)(r+s)}{s(r+\lambda_n+s)}\, .
\label{sepvar.2}
\end{equation}
Substituting this result in Eq. (\ref{sepvar.1}) gives
\begin{equation}
\tilde{p}(x,s)=\frac{(r+s)}{s}\, \sum_{\lambda_n} \frac{f_{\lambda_n}(0)}{r+\lambda_n+s}\,
\left[ e^{-\frac{1}{2D}\, U(x)} \psi_{\lambda_n}(x)\right]\, .
\label{sepvar.3}
\end{equation}
Now, setting $r=0$ in Eq. (\ref{sepvar.3}) gives
\begin{equation}
\tilde{p}_0(x,s)= \sum_{\lambda_n} \frac{f_{\lambda_n}(0)}{\lambda_n+s}\,
\left[ e^{-\frac{1}{2D}\, U(x)} \psi_{\lambda_n}(x)\right]\, .
\label{sepvar.4}
\end{equation}
Comparing Eqs. (\ref{sepvar.3}) and (\ref{sepvar.4}) it follows that
\begin{equation}
\tilde{p}(x,s)= \frac{(r+s)}{s}\, \tilde{p}_0(x,r+s)\, ,
\label{sepvar.5}
\end{equation}
which then fully reproduces the renewal result in Eq. (\ref{laplace.1}).


\section*{References}
\begin{thebibliography}{20}

\bibitem{Stanislavsky00} A. A. Stanislavsky, {\em Memory
  effects and macroscopic manifestation of randomness},  Phys. Rev.
E {\bf 61}, 4752 (2000).

\bibitem{BB2008}
 S.  Burov and E. Barkai, {\em Fractional Langevin equation:
 Overdamped, underdamped, and critical behaviors}, Phys. Rev.
 E {\bf 78}, 031112 (2008).
    
\bibitem{MSVV2013}
C. Maes, S. Safaverdi, P. Visco, F. Van Wijland, {\em Fluctuation-response relations for nonequilibrium diffusions
with memory}, Phys.  Rev. E {\bf 87}), 022125 (2013).

\bibitem{MK2000} R. Metzler and J.  Klafter, The random walk's guide to anomalous diffusion: a fractional dynamics approach,
{\em Phys. Rep.} {\bf 339}, 1 (2000).

\bibitem{EL2018}L. R. Evangelista and E. K. Lenzi, {\em Fractional Diffusion Equations and Anomalous Diffusion} (Cambridge University Press, Cambridge, 2018).

\bibitem{B1992}
J. P. Bouchaud, {\em Weak ergodicity breaking and aging in disordered systems}, J. Phys. I France {\bf 2}, 1705 (1992).

\bibitem{BB2005} G. Bel and E. Barkai, {\em Weak ergodicity breaking in the continuous-time random walk}, Phys. Rev. Lett. {\bf 94}, 240602 (2005).

\bibitem{B2016} A. A. Budini, {\em Weak ergodicity breaking induced by global memory effects},
Phys. Rev. E {\bf 94}, 022108 (2016).

\bibitem{FCBGM2017}
A. Falc\'on-Cort\'es, D.  Boyer, L. Giuggioli and S. N. Majumdar,  {\em Localization transition induced by learning in random searches}, Phys. Rev. Lett. {\bf 119}, 140603 (2017).

\bibitem{BEM2017} D. Boyer, M. R. Evans, and S. N. Majumdar, {\em Long time scaling behaviour for 
diffusion with resetting and memory}, J. Stat. Mech. 023208 (2017).

\bibitem{EM2011} M. R. Evans and S. N. Majumdar, {\em Diffusion with stochastic resetting}, Phys. Rev. 
Lett. {\bf 106}, 160601 (2011).


\bibitem{BS2014} D. Boyer and C. Solis-Salas, {\em Random walks with preferential relocations to places 
visited in the past and their application to biology}, Phys. Rev. Lett. {\bf 112}, 240601 (2014).

\bibitem{BR2014}
D. Boyer and  J. C. R. Romo-Cruz,
    {\em Solvable random-walk model with memory and its relations with Markovian models of anomalous diffusion},
Phys. Rev. E {\bf 90}, 042136 (2014).

\bibitem{MPCM2019} A Mas\'o-Puigdellosas, D.  Campos and V. M\'endez, {\em Anomalous diffusion in random-walks with memory-induced relocations}, Frontiers in Physics {\bf 7}, 112 (2019).

\bibitem{MUB2019} C. Mailler  and G. Uribe-Bravo,  {\em  Random walks with preferential relocations and fading memory:
    a study through random recursive trees},  J. Stat. Mech. 093206 (2019).
  
\bibitem{BMa2023} E-S. Boci and C. Mailler,  {\em Large deviation principle for a stochastic process with random
    reinforced relocations}, J. Stat. Mech. 083206 (2023).

\bibitem{BMa2024} E-S. Boci and C. Mailler, {\em Central limit theorems for the monkey walk with steep memory kernel}, arXiv preprint arXiv:2409.02861 (2024).


\bibitem{BM2024} D. Boyer and S. N. Majumdar, {\em Power-law relaxation of a confined diffusing particle subject to 
resetting with memory}, J. Stat. Mech. 073206 (2024).

  
\bibitem{EMS2020} M. R. Evans, S. N. Majumdar, G. Schehr,
{\em Stochastic resetting and applications}, J. Phys. A: Math. Theor. {\bf 53}, 193001 (2020).



\bibitem{P2015} A. Pal, {\em Diffusion in a potential landscape with stochastic resetting}, Phys. Rev. E {\bf 91}, 012113 (2015).

\bibitem{Risken} H. Risken, {\em The Fokker-Planck Equation: Methods of Solutions and Applications} (Springer-Verlag, Berlin, 1996).

\bibitem{AS_book} M. Abramowitz and I. A. Stegun, {\em Handbook of Mathematical Functions: with
Formula, Graphs, and Mathematical Tables} (Dover, New York, 1965).

\bibitem{CTDL1977} 
C. Cohen-Tannoudji, B. Diu and F. Lalo\"e, {M\'ecanique Quantique I} (Hermann, Paris, 1977).

\bibitem{EM14}
 M. R. Evans and S. N. Majumdar {\em Diffusion with resetting in arbitrary spatial dimension},  J. Phys. A: Math. Theor. {\bf 47} 285001 (2014).





\end{thebibliography}
\end{document}